\documentclass[letters,fleqn,usenatbib]{mnras}

\usepackage{newtxtext,newtxmath}

\usepackage[T1]{fontenc}
\usepackage{aecompl}

\usepackage{graphicx}	
\usepackage{subfigure}	
\usepackage{amsmath}	
\usepackage{amssymb}	
\usepackage{textcomp}
\usepackage[all,cmtip]{xy}
\usepackage{tikz}
\usetikzlibrary{positioning,arrows.meta,spy}

\title[The origin of the LMC bar]{The origin of the LMC stellar bar: clues from the SFH of the bar and inner disk.}

\author[L. Monteagudo et al.]{
L.\, Monteagudo,$^{1,2}$\thanks{E-mail: laram@iac.es}
C.\, Gallart,$^{1,2}$
M.\, Monelli,$^{1,2}$
E.\,J. Bernard,$^{3}$
P.\,B.\, Stetson$^{4}$
\\
$^{1}$Instituto de Astrof\'isica de Canarias (IAC), Calle V\'ia L\'actea s/n, E-38205 La Laguna, Tenerife; Spain\\
$^{2}$Departamento de Astrof\'isica, Universidad de La Laguna (ULL), E-38206 La Laguna, Tenerife; Spain\\
$^{3}$Universit\'e C\^ote d'Azur, OCA, CNRS, Lagrange, France\\
$^{4}$Herzberg Astronomy and Astrophysics, National Research Council Canada, 5071 West Saanich Road, Victoria, BC V9E 2E7, Canada\\
}

\date{Accepted XXX. Received YYY; in original form ZZZ}

\pubyear{2016}

\begin{document}
\label{firstpage}
\pagerange{\pageref{firstpage}--\pageref{lastpage}}
\maketitle

\begin{abstract}

We discuss the origin of the LMC stellar bar by comparing the star formation 
histories (SFH) obtained from deep color-magnitude diagrams (CMDs) in the bar and in 
a number of fields in different directions within the inner disk. The CMDs, reaching 
the oldest main sequence turnoffs in these very crowded fields, have been obtained 
with VIMOS on the VLT in service mode, under very good seeing conditions. We show 
that the SFHs of all fields share the same patterns, with consistent variations of 
the star formation rate as a function of time in all of them. We therefore conclude 
that no specific event of star formation can be identified with the formation of the LMC bar,  which instead likely formed from a redistribution of disk material that occurred when the LMC disk became bar unstable, and shared a common SFH with the inner disk thereafter.  
The strong similarity between the SFH of the center and edge of the bar rules out significant spatial variations of the SFH across the bar, which are predicted by scenarios of classic bar formation through buckling mechanisms.  

\end{abstract}




\textbf{\textbf{}}\section{Introduction}\label{sec:introduccion}

The Large Magellanic Cloud is the prototype of a whole class of 
galaxies, the Barred Magellanic Spirals (SBm), characterized by 
the presence of an optically visible stellar bar, coincident or 
not with the dynamical center of the 
galaxy, a single spiral arm emanating from an end of the bar, and 
often, a large star forming region at one end of the bar \citep{Vaucouleurs72}. 

The true nature of the bars in these late-type galaxies is the subject of controversy.  While  in early-type spirals the barred optical morphology is 
also evident in both the distribution and kinematics of the neutral HI gas, this is 
not always the case in SBm; in many examples, the bar seems to have a modest 
effect (if any) on the gas kinematics toward the center of the 
galaxy \citep{Wilcots08}.

The existence of the LMC as a very nearby representative of the SBm class offers 
an excellent opportunity to gain insight on the origin and evolution of these barred 
structures.  In the LMC, a stellar bar is clearly visible in near-IR maps and stellar 
density contours \citep[e.g.][]{marel01, cioni00} but is not apparent in the HI gas 
disk \citep{Staveley03, Kim98} and is not the site of current star formation as shown
by H$\alpha$ images \citep{Kim99}. There has been a fair amount of discussion regarding 
its three-dimensional structure \citep[e.g.][]{Zhao00, Zaritsky04} and its location with respect to 
the stellar disk \citep{Subramaniam03, Nikolaev04, Lah05, Koerwer09,Subramaniam09}. 


A few studies have addressed the star-formation history (SFH) of the LMC bar and nearby 
fields. The high stellar density in the center of the bar, however, has made it difficult 
to obtain color-magnitude diagrams (CMD) reaching the oldest main sequence turnoffs 
(oMSTO) with ground-based telescopes, necessary for a reliable determination of the SFH for all ages.
The WFPC2 on HST produced the first very deep 
CMDs of fields in the LMC bar \citep{elson97, Holtzman99, Olsen99, smeckerhane02, 
Weisz13}, which were populated enough to lead to reliable SFHs; but these were for 
small portions of the bar and were thus potentially affected by local fluctuations in
the stellar populations. Deep and well populated CMDs of the inner LMC disk immediately 
surrounding the bar region were even more challenging, because a single WFPC2 field 
typically produced very sparse CMDs. To address this issue, some programs observed 
mosaics of several WFPC2 fields in the inner LMC disk in different directions from 
the bar center, and produced CMDs as populated as those in the bar \citep{smeckerhane02}. Comprehensive coverage of the LMC bar and inner disk using the HST, however, remains unfeasible. 

For this reason, we have adopted an alternative approach using ground-based observations, taken in excellent seeing conditions with the VLT in service mode. This strategy produces CMDs reaching the oMSTO even in the center of the LMC bar and allows us to study representative portions of the inner LMC for the first time, leading to sound conclusions on their entire SFH. As part of a larger project devoted to an in-depth study of the central LMC SFH and its spatial variations, in this paper we show how the striking similarity of the SFHs of a number of bar and disk fields puts strong constraints on the formation of the LMC bar. This has obvious importance for addressing the long-standing issue regarding the nature of the LMC bar and its possibly differentiated formation and/or evolution. 

\section{Observations and data reduction}\label{sec:ovservations}

\begin{figure}
	\includegraphics[width=\columnwidth]{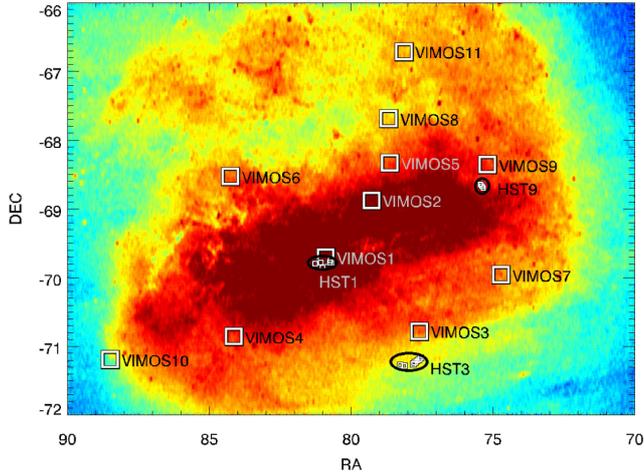}
	\caption{Stellar density map of the central region of the LMC, based on 
the photometry of the Gaia Data Release 1. The location of our VIMOS fields and the archival HST pointings are overlaid.}
	\label{fig:vimos_1}
\end{figure}

We have used the VIsible Multi-Object Spectrograph (VIMOS) on the 
Very Large Telescope (VLT) to obtain deep B and R images in the 
central part of the LMC. The camera has four CCDs, each with a 7' x 8' 
field of view. We observed a total of eleven fields in order to have
a significant sampling of the LMC's innermost region. Two fields probe 
the LMC bar, and the remaining nine fields are distributed in a ring in 
the inner disk and northern arm within a distance of R<3.2$^{\circ}$ 
from the bar center. Fig~\ref{fig:vimos_1} shows the location of 
these VIMOS fields, superimposed on a stellar density map of the LMC 
based on data from the Gaia DR1 \citep{GaiaDR1}.

The observations were designed to reach the oMSTO in the CMD, necessary to 
obtain a reliable full life-time SFH. They were taken in service mode to ensure the good 
seeing $\simeq { 0.6''-0.8''}$ necessary to resolve stars down to a faint 
magnitude limit in these very crowded areas. 

We obtained the photometry using DAOPHOT IV and ALLFRAME 
\citep{stetson87,stetson94}. Each chip of each image was reduced 
independently. Photometric calibration is based on a number of standard fields 
observed during a photometric campaign with the CTIO Blanco Telescope, with 
the MOSAIC camera, on January 15, 2010. In this period we observed the 
same eleven fields and a number of standard fields, which were selected 
because of the large number of standard stars available in the 
database of P.B. Stetson.\footnote{http://www3.cadc-ccda.hia-iha.nrc-cnrc.gc.ca/community/STETSON/standards/}
Finally, a large number of artificial-star tests were performed in 
each frame following the procedure described in \citet{gallart99}. 
These are used both to derive completeness factors and to model 
photometric errors in the synthetic CMD. 

To facilitate the comparison of our SFHs with previous work, we have 
supplemented our ground-based data with deep archival HST imaging. These 
HST data are based on WFPC2 images from several programs: GO7382 and GO8576  (P.I. 
Smecker-Hane), GO7306 (P.I. Cook) and GO6229 (P.I. Trauger). We used 
photometry and artificial-star tests ($\sim 1.2 \times 10^{5}$ per 
field) taken from the Local Group Stellar Photometry 
Archive\footnote{http://astronomy.nmsu.edu/holtz/archival/html/lg.html} 
\citep[LGSPA:][]{holtzman06}. Fig.~\ref{fig:vimos_1} shows that the
mosaics of disk WFPC2 fields and several WFPC2 fields in the bar are clustered in three small regions of the LMC, which are very close to three of our fields, namely VIMOS1, VIMOS3 and VIMOS9. We have combined the HST observations in three CMDs that we will call HST1, HST3 and HST9 to indicate their closeness to a 
VIMOS field.

\begin{figure}
	\includegraphics[width=\columnwidth,height=12cm]{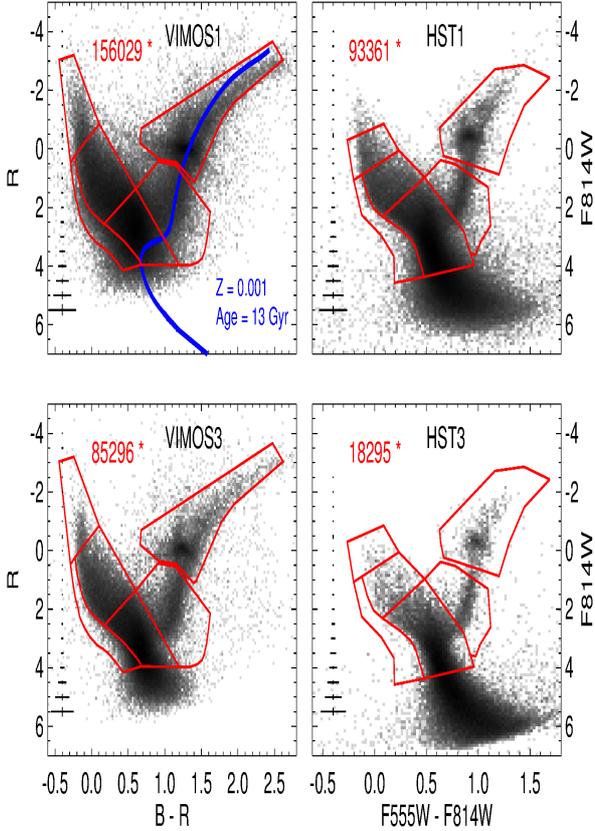}
	\caption{ {\bf Left panels}: CMDs of two of the VIMOS fields which have a nearby mosaic of WFPC2 data. The blue line in the upper panel is a 13 Gyr old BaSTI isochrone with Z=0.001; it highlights that even in the most central, crowded field our photometry is deep enough to reach the oMSTO. {\bf Right panels}: CMDs of the WFPC2 fields spatially located next to VIMOS fields whose CMDs are displayed in the left panels. Red lines delimit the bundles containing the stars we used for the SFH calculation. The numbers of stars inside the bundles are labeled. Errorbars indicating photometric uncertainties as a function of magnitude are shown in each panel.}
	\label{fig:dcm}
\end{figure}

Fig.~\ref{fig:dcm} shows a sample of the VIMOS (left panels) and HST (right) CMDs, for the central bar field (VIMOS1 and HST1 on the top) and one representative disk field (VIMOS3 and HST3). In the case of the bar fields, CMDs from a single WFPC2 pointing contain a number of stars that is sufficient for a robust determination of the 
SFH \citep{Holtzman99, Olsen99, smeckerhane02}. For the disk, HST GO programs 7382 
and 8576 \citep[see][]{smeckerhane02} mosaicked 6--10 WFPC2 pointings to 
obtain CMDs with a number of stars comparable to that of a bar field. These HST CMDs 
are deeper than the VIMOS ones, but the latter have the advantage of containing a 
much larger number of stars (see Fig.~\ref{fig:dcm}).

An old isochrone has been superimposed on the ground-based CMD of the central bar field (top left panel of Fig.~\ref{fig:dcm}) to show that our goal to 
reach the oMSTO in the CMD was achieved in our most crowded field. To 
our knowledge, these are the only data taken from the ground with 
photometry deep enough to reach the oMSTO in the center of the LMC bar. 
Fig.~\ref{fig:dcm} also shows the {\it bundles}, or areas of the CMD that have been used to derive the SFH through comparison of the distribution of stars in the 
observed and synthetic CMDs (see Section~\ref{sec:method}). The number of stars 
inside the bundles has been labeled in each CMD. Note that, in spite 
of the fact that several WFPC2 fields have been combined to build the 
CMDs shown in Fig.~\ref{fig:dcm}, the number of stars in the 
VIMOS CMDs that are relevant for the SFH derivation is many times 
greater than in the WFPC2 CMDs.  

\section{The star formation history}\label{sec:sfh}

\subsection{SFH Derivation}\label{sec:method}

The SFH calculations for both VIMOS and WFPC2 data were carried out 
using the CMD-fitting technique, in a way very similar to that described in 
\citet{aparicio09} and \citet{meschin14}. We used 
IAC-star\footnote{http://iac-star.iac.es} \citep{aparicio04} to compute a synthetic CMD with a constant star formation rate (SFR(t)) between 13.5 and $\simeq$ 0.03 Gyr ago; 5 x 10$^{7}$ stars in the whole age range are uniformly distributed between Z = 0.0001 and 0.02 (-2.3 $\leq$ [Fe/H] $\leq$ 0.004, assuming Z$_\odot$=0.0198). The BaSTI stellar evolution library \citep[][solar scaled, overshooting set]{Pietrinferni04} has been adopted.  For the initial mass function (IMF) and the  binary star distribution function $\beta\textit{(f,q)}$ we used the same values as in \citet{meschin14}: a binary fraction $\textit{f}$ = 0.4 and a mass ratio distribution $\textit{q}$ > 0.5. The IMF was taken from \citet{Kroupa02} and 
is given by $\textit{N(m)dm = m$^{-\alpha}$dm}$, where $\alpha$=1.3 for stars with mass 0.1 $\leqslant$ $\textit{m/M}$$\odot$ $\leqslant$ 0.5, and $\alpha$=2.3 for 
0.5$\leqslant$ $\textit{m/M}$$\odot$ $\leqslant$ 100.

The incompleteness and photometric uncertainties due to the observational 
effects have been simulated in the synthetic CMD for each VIMOS and 
HST pointing based on the results of the corresponding artificial-star 
tests. In the the case of the two bar fields, we identified areas with reddening larger than average and removed the corresponding stars from the CMD used to derive the SFH (approximately one third of the stars were removed for this reason).  We verified, however, that the changes in the SFH when including these regions are minimal. No significant effect due to differential reddening could be noticed in the CMD of the disk fields. To obtain the SFHs, we used a new algorithm developed in 
Python by one of us (EJB) \citep[see][for some details 
on the algorithm]{bernard15}. The same method was applied to all the 
VIMOS fields and three HST groups. 

\subsection{Comparison of VIMOS and HST SFHs}\label{sec:comparison}

\begin{figure}
 \includegraphics[width=\columnwidth,height=5cm]{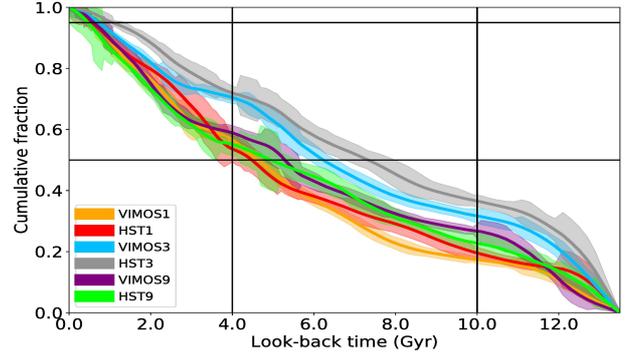}
 \caption{Comparison of the cumulative SFHs obtained from the VIMOS and WFPC2 data, for the three LMC regions for which the two data sources are available. The horizontal lines indicate mass fractions corresponding to 50\% and 95\% of the total accumulated mass. Vertical lines indicate the approximate ages that separate the main star forming episodes mentioned in the text.}
 \label{fig:sfh_cum}
 \end{figure}

In this section we compare the SFH of the two VIMOS fields located next
to HST ones to show that the SFHs derived from 
the VIMOS data are compatible with those obtained from the deeper HST data. 
The results, in the form of cumulative SFHs, are displayed in 
Fig.~\ref{fig:sfh_cum}. It can be seen that the SFHs for each VIMOS 
and corresponding WFPC2 field are basically identical within the errors, 
while there are noticeable differences between fields. The largest 
difference in photometric depth between the VIMOS and corresponding 
WFPC2 field occurs for the bar field (\#1), for which the VIMOS CMD 
reaches just about half a magnitude below the oMSTO. The results in 
Fig.~\ref{fig:sfh_cum}, however, show that the features in the SFH are 
equally recovered from both CMDs. The much larger number of stars in the 
less deep VIMOS CMD and, particularly, the fact that it does reach the 
oMSTO, are likely the key reasons for this.

\subsection{The SFH of the LMC bar and inner disk 
from VIMOS data}\label{sec:vimos}

We obtained SFHs for each individual VIMOS field and compared them. The detailed results on individual fields, including both SFR(t) and age-metallicity relations, will be presented in a future paper (Monteagudo et al. 2018, in preparation). Disk fields \#3, 
4, 5, 6, 7, 8 and 10 have very similar SFHs, characterized by relatively smooth 
variations of the SFR(t) with respect to a mean value, over the 
whole lifetime of the galaxy, resulting in a cumulative SFR(t) close to a constant value. Therefore, for clarity in  Fig.~\ref{fig:sfh_cum} we only show the SFH of field \#3. The SFH of the two bar fields indicates a stellar population younger overall, while that of disk fields \#9 and 11 (see SFH for field \#9 in Fig.~\ref{fig:sfh_cum}) is somewhat intermediate between the remaining disk fields and the bar fields.  Since these two fields are located in the North LMC arm, we will exclude them from the upcoming analysis, focused on the comparison of the SFHs of the bar and inner disk. However, including them in the analysis would not change the conclusions of the paper.  

\begin{figure}
	\includegraphics[width=\columnwidth]{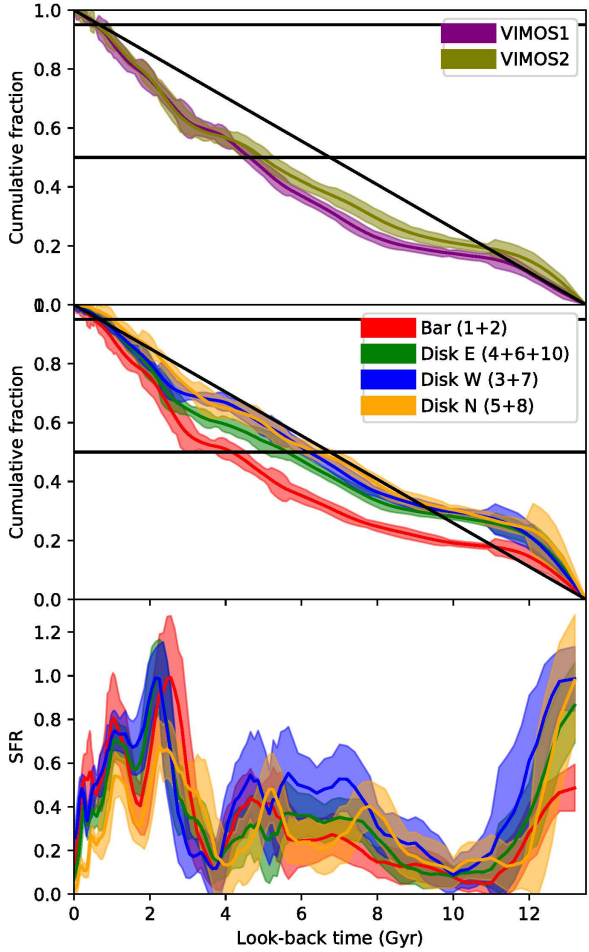}
	\caption{{\bf Top panel:} Cumulative SFH for the LMC bar fields, \#1 and 
\#2. The horizontal lines indicate mass fractions corresponding to 50\% and 
95\% of the total accumulated mass. {\bf Middle and lower panels}: SFHs (in 
cumulative and time resolved form, respectively) for the bar and disk fields, 
grouped as indicated in the text.}
	\label{fig:sfh}
\end{figure}

In this paper, we are interested in exploring possible variations of the SFH within 
the bar, and between the bar and the surrounding inner disk fields. Therefore, for the 
purposes of the current paper, we have combined the 
CMDs of disk fields with similar SFHs into three CMDs that we will 
consider representative of the stellar populations of the inner disk 
in the N, E, and SW directions.  DiskE will correspond to fields with 
$\alpha_{2000}$ $\geqslant$ 05:30:00 (VIMOS4+6+10), DiskSSW to fields 
with $\alpha_{2000}$ $\leqslant$ 05:10:00 and $\delta_{2000}$ $\geqslant$ -68:00:00 (VIMOS3+7) and DiskN to those with $\alpha_{2000}$ $\backsimeq$ 05:14:00 (VIMOS5+8). For the bar fields (\#1 and \#2) we have computed individual SFHs, which are represented in the upper panel of Fig.~\ref{fig:sfh} in cumulative form. Note that both are almost identical within the errors (with the SFH for the central bar field marginally younger than that of the field in the NW extreme of the bar) indicating a basically common SFH for the whole bar. We have thus combined both bar fields for further comparison with the disk. These comparisons are shown in the middle 
and lower panel of Fig.~\ref{fig:sfh} in cumulative and time resolved form 
respectively. 
In all figures, the SFHs are represented with their corresponding 
uncertainties, estimated following the prescriptions of \citet{hidalgo11}. 

The lower panel of Fig.~\ref{fig:sfh} shows that the SFR(t) of the bar and combined 
disk fields presents common features and consistent trends. All are characterized by 
three main periods of star formation separated by short gaps of almost negligible star 
formation activity. We find an early star formation episode (\textit{old star-forming epoch}, 
O$_{SFE}$) common to all fields and lasting $\simeq$ 3.5 Gyr. A second period of enhanced SFR(t) 
(\textit{intermediate star-forming epoch}, I$_{SFE}$) is found between 10 and 4 Gyr 
ago. Finally, the most recent period (\textit{young star-forming epoch}, Y$_{SFE}$) 
began $\sim$ 4 Gyr ago. Within each period, there are variations in the intensity of the SFR(t) which are only slightly different in detail from field to field. In particular, it is interesting to note that, within the young star-forming epoch, which is the one for which we can be most confident on the details of SFR(t), there are variations that are totally consistent among all fields. Three 
peaks of star formation activity are observed at $\simeq$ 2.5, 1.0 and 0.5 Gyr ago, 
while star formation appears very much reduced at the present time.

What is different between bar and disk fields is the relative number 
of stars formed in the three main epochs of star formation. In the first three lines 
of Tab.~\ref{tab:integrales} we indicate the fraction of stars formed in each of 
them, for the combined bar fields and disk fields. The fraction of 
stars formed in the O$_{SFE}$ is lower in the bar field than in the disk fields, and 
the contrary is true for the Y$_{SFE}$, while the fraction formed at intermediate 
ages is similar in all fields. This leads to a ratio Y/O and Y/(I+O) about a factor 
of two larger in the bar compared to the disk.  These differences are reflected in 
the respective cumulative mass fractions, displayed in the middle panel of 
Fig.~\ref{fig:sfh}. In this figure, the two horizontal lines indicate the 50 
and 95 mass percentiles. The ages at which these percentiles are reached in each 
field are listed in Tab.~\ref{tab:integrales}. They indicate that the disk formed half of its mass between 1.25 and 2 Gyr earlier than the bar.

\subsection{Discussion: the origin of the LMC bar}\label{sec:discussion}

The highly detailed SFHs that we have derived for a number of regions 
covering representative portions of the LMC bar and inner disk allow us to provide 
important constraints on the nature of the LMC bar. In the previous section we
have shown that the SFHs of the bar and disk fields closely share the same features, 
and thus, {\it no event of star formation can be identified with the formation of 
the LMC bar.} This conclusion is different from that reached in previous studies 
\citep{elson97, smeckerhane02}. In particular, \citet{smeckerhane02} identified a 
4--6 Gyr star formation episode with the formation of the LMC bar. The significantly 
larger fields, covering different positions in the LMC inner regions, and the more 
sophisticated analysis technique \citep[][simply modeled the main 
sequence luminosity function to derive the SFH]{smeckerhane02}, makes us confident 
that our conclusion is robust. It implies that the bar likely 
formed from a redistribution of disk material that occurred when the disk became bar 
unstable, and shared a common SFH with the inner disk thereafter. 

The fact that the Y$_{SFE}$ has been somewhat more intense in the bar (and in 
its innermost region) than in the inner disk might be a consequence of younger, 
colder material and gas being preferentially funneled to the center of the galaxy by 
the non-axisymmetric potential. However, it may also simply be a continuation of the 
gradient seen in the outer disk \citep{gallart08,meschin14}, and common in dwarf 
irregular galaxies \citep[e.g.][]{bernard07, stinson09, hidalgo13} in the sense that 
younger populations are concentrated toward the central parts of the galaxies. 

The fact that the two bar fields, one located in its very center and the other on 
its northern rim, also share a closely similar SFH allows us to put further 
constraints on the characteristics of the LMC bar. \citet{friedli95} showed that 
the formation of a strong bar in a typical Sc disk induces a significant starburst 
in the bar and in the galactic center. The lack of an excess of H$_\alpha$ 
emission in the bar region \citep{Kim99} indicates that such a starburst is not currently
ongoing and our SFH results allow us to reach the same conclusion for the rest of 
the galaxy's lifetime. This kind of predicted variations of the SFH within the bar 
caused by bar formation and buckling have been recently observed in the SBb 
galaxy NGC$\,$6032, where it was also observed that the SFH of the outer bar was 
similar to that of the disk~\citep{Perez2017}. The basically identical SFH across the LMC bar points to the absence of these buckling mechanisms characteristic of classical bars 
in more massive galaxies. A more comprehensive mapping of the SFH across the whole 
LMC bar is necessary to confirm this point, which places important constraints on 
the formation of bars in low mass galaxies, particularly of the Magellanic type. 


\begin{table}
	\centering
	\caption{Derived values from the SFH}
	\label{tab:integrales}
	\begin{tabular}{lcccc} 
		\hline
		\hline
		& Bar & Disk E & Disk SW & Disk N \\
		\hline
		$Y_{SFE}$ & 0.50$\pm$0.03 & 0.40$\pm$0.02 & 0.33$\pm$0.02 & 0.35$\pm$0.02\\
		$I_{SFE}$ & 0.31$\pm$0.03 & 0.31$\pm$0.02 & 0.37$\pm$0.03 & 0.34$\pm$0.03\\
		$O_{SFE}$ & 0.19$\pm$0.02 & 0.28$\pm$0.03 & 0.30$\pm$0.03 & 0.30$\pm$0.05\\ 
		(Y/O)$_{SFE}$ & 2.6$\pm$0.3 & 1.45$\pm$0.18 & 1.12$\pm$0.13 & 1.16$\pm$0.21 \\
		(Y/(I+O))$_{SFE}$ & 1.00$\pm$0.09 & 0.68$\pm$0.05 & 0.49$\pm$0.04 & 0.55$\pm$0.07 \\ 
		T$_{50\% }$ (Gyrs) & 4.25 & 5.5 & 6.25 & 6.25 \\ 
		T$_{95\% }$ (Gyrs) & 0.5 & 0.75 & 0.75 & 1.00 \\ 
		\hline
	\end{tabular}
\end{table}


\section*{Acknowledgements}

We thank I. P\'erez, I. Mart\'\i nez-Valpuesta and T. Ru\'\i z-Lara for useful discussions. This research is based on observations made with the ESO VLT at the La Silla Paranal Observatory under programme ID 084.B-1124, and in observations made with the NASA/ESA Hubble Space Telescope, obtained from the data archive at the Space Telescope Science Institute. STScI is operated by AURA under NASA contract NAS 5-26555. This work has been supported by the Spanish Ministry of Economy and Competitiveness (MINECO) under grant AYA2014-56795-P. EJB acknowledge support from the CNES postdoctoral fellowship program.




\bibliographystyle{mnras}
\bibliography{ref} 

\begin{thebibliography}{}
\makeatletter
\relax
\def\mn@urlcharsother{\let\do\@makeother \do\$\do\&\do\#\do\^\do\_\do\%\do\~}
\def\mn@doi{\begingroup\mn@urlcharsother \@ifnextchar [ {\mn@doi@}
  {\mn@doi@[]}}
\def\mn@doi@[#1]#2{\def\@tempa{#1}\ifx\@tempa\@empty \href
  {http://dx.doi.org/#2} {doi:#2}\else \href {http://dx.doi.org/#2} {#1}\fi
  \endgroup}
\def\mn@eprint#1#2{\mn@eprint@#1:#2::\@nil}
\def\mn@eprint@arXiv#1{\href {http://arxiv.org/abs/#1} {{\tt arXiv:#1}}}
\def\mn@eprint@dblp#1{\href {http://dblp.uni-trier.de/rec/bibtex/#1.xml}
  {dblp:#1}}
\def\mn@eprint@#1:#2:#3:#4\@nil{\def\@tempa {#1}\def\@tempb {#2}\def\@tempc
  {#3}\ifx \@tempc \@empty \let \@tempc \@tempb \let \@tempb \@tempa \fi \ifx
  \@tempb \@empty \def\@tempb {arXiv}\fi \@ifundefined
  {mn@eprint@\@tempb}{\@tempb:\@tempc}{\expandafter \expandafter \csname
  mn@eprint@\@tempb\endcsname \expandafter{\@tempc}}}

\bibitem[\protect\citeauthoryear{{Aparicio} \& {Gallart}}{{Aparicio} \&
  {Gallart}}{2004}]{aparicio04}
{Aparicio} A.,  {Gallart} C.,  2004, \mn@doi [\aj] {10.1086/382836}, \href
  {http://adsabs.harvard.edu/abs/2004AJ....128.1465A} {128, 1465}

\bibitem[\protect\citeauthoryear{{Aparicio} \& {Hidalgo}}{{Aparicio} \&
  {Hidalgo}}{2009}]{aparicio09}
{Aparicio} A.,  {Hidalgo} S.~L.,  2009, \mn@doi [\aj]
  {10.1088/0004-6256/138/2/558}, \href
  {http://adsabs.harvard.edu/abs/2009AJ....138..558A} {138, 558}

\bibitem[\protect\citeauthoryear{{Bernard}, {Aparicio}, {Gallart},
  {Padilla-Torres}  \& {Panniello}}{{Bernard} et~al.}{2007}]{bernard07}
{Bernard} E.~J.,  {Aparicio} A.,  {Gallart} C.,  {Padilla-Torres} C.~P.,
  {Panniello} M.,  2007, \mn@doi [\aj] {10.1086/520805}, \href
  {http://adsabs.harvard.edu/abs/2007AJ....134.1124B} {134, 1124}

\bibitem[\protect\citeauthoryear{{Bernard}, {Ferguson}, {Chapman}, {Ibata},
  {Irwin}, {Lewis}  \& {McConnachie}}{{Bernard} et~al.}{2015}]{bernard15}
{Bernard} E.~J.,  {Ferguson} A.~M.~N.,  {Chapman} S.~C.,  {Ibata} R.~A.,
  {Irwin} M.~J.,  {Lewis} G.~F.,   {McConnachie} A.~W.,  2015, \mn@doi [\mnras]
  {10.1093/mnrasl/slv116}, \href
  {http://adsabs.harvard.edu/abs/2015MNRAS.453L.113B} {453, L113}

\bibitem[\protect\citeauthoryear{{Cioni}, {Habing}  \& {Israel}}{{Cioni}
  et~al.}{2000}]{cioni00}
{Cioni} M.-R.~L.,  {Habing} H.~J.,   {Israel} F.~P.,  2000, \aap, \href
  {http://adsabs.harvard.edu/abs/2000A%26A...358L...9C} {358, L9}

\bibitem[\protect\citeauthoryear{{Elson}, {Gilmore}  \& {Santiago}}{{Elson}
  et~al.}{1997}]{elson97}
{Elson} R.~A.~W.,  {Gilmore} G.~F.,   {Santiago} B.~X.,  1997, \mn@doi [\mnras]
  {10.1093/mnras/289.1.157}, \href
  {http://adsabs.harvard.edu/abs/1997MNRAS.289..157E} {289, 157}

\bibitem[\protect\citeauthoryear{{Friedli} \& {Benz}}{{Friedli} \&
  {Benz}}{1995}]{friedli95}
{Friedli} D.,  {Benz} W.,  1995, \aap, \href
  {http://adsabs.harvard.edu/abs/1995A%26A...301..649F} {301, 649}

\bibitem[\protect\citeauthoryear{{Gaia Collaboration} et~al.,}{{Gaia
  Collaboration} et~al.}{2016}]{GaiaDR1}
{Gaia Collaboration} et~al., 2016, \mn@doi [\aap]
  {10.1051/0004-6361/201629512}, \href
  {http://adsabs.harvard.edu/abs/2016A%26A...595A...2G} {595, A2}

\bibitem[\protect\citeauthoryear{{Gallart}, {Freedman}, {Aparicio}, {Bertelli}
  \& {Chiosi}}{{Gallart} et~al.}{1999}]{gallart99}
{Gallart} C.,  {Freedman} W.~L.,  {Aparicio} A.,  {Bertelli} G.,   {Chiosi} C.,
   1999, \mn@doi [\aj] {10.1086/301078}, \href
  {http://adsabs.harvard.edu/abs/1999AJ....118.2245G} {118, 2245}

\bibitem[\protect\citeauthoryear{{Gallart}, {Stetson}, {Meschin}, {Pont}  \&
  {Hardy}}{{Gallart} et~al.}{2008}]{gallart08}
{Gallart} C.,  {Stetson} P.~B.,  {Meschin} I.~P.,  {Pont} F.,   {Hardy} E.,
  2008, \mn@doi [\apjl] {10.1086/590552}, \href
  {http://adsabs.harvard.edu/abs/2008ApJ...682L..89G} {682, L89}

\bibitem[\protect\citeauthoryear{{Hidalgo} et~al.,}{{Hidalgo}
  et~al.}{2011}]{hidalgo11}
{Hidalgo} S.~L.,  et~al., 2011, \mn@doi [\apj] {10.1088/0004-637X/730/1/14},
  \href {http://adsabs.harvard.edu/abs/2011ApJ...730...14H} {730, 14}

\bibitem[\protect\citeauthoryear{{Hidalgo} et~al.,}{{Hidalgo}
  et~al.}{2013}]{hidalgo13}
{Hidalgo} S.~L.,  et~al., 2013, \mn@doi [\apj] {10.1088/0004-637X/778/2/103},
  \href {http://adsabs.harvard.edu/abs/2013ApJ...778..103H} {778, 103}

\bibitem[\protect\citeauthoryear{{Holtzman} et~al.,}{{Holtzman}
  et~al.}{1999}]{Holtzman99}
{Holtzman} J.~A.,  et~al., 1999, \mn@doi [\aj] {10.1086/301097}, \href
  {http://adsabs.harvard.edu/abs/1999AJ....118.2262H} {118, 2262}

\bibitem[\protect\citeauthoryear{{Holtzman}, {Afonso}  \& {Dolphin}}{{Holtzman}
  et~al.}{2006}]{holtzman06}
{Holtzman} J.~A.,  {Afonso} C.,   {Dolphin} A.,  2006, \mn@doi [\apjs]
  {10.1086/507074}, \href {http://adsabs.harvard.edu/abs/2006ApJS..166..534H}
  {166, 534}

\bibitem[\protect\citeauthoryear{{Kim}, {Staveley-Smith}, {Dopita}, {Freeman},
  {Sault}, {Kesteven}  \& {McConnell}}{{Kim} et~al.}{1998}]{Kim98}
{Kim} S.,  {Staveley-Smith} L.,  {Dopita} M.~A.,  {Freeman} K.~C.,  {Sault}
  R.~J.,  {Kesteven} M.~J.,   {McConnell} D.,  1998, \mn@doi [\apj]
  {10.1086/306030}, \href {http://adsabs.harvard.edu/abs/1998ApJ...503..674K}
  {503, 674}

\bibitem[\protect\citeauthoryear{{Kim}, {Dopita}, {Staveley-Smith}  \&
  {Bessell}}{{Kim} et~al.}{1999}]{Kim99}
{Kim} S.,  {Dopita} M.~A.,  {Staveley-Smith} L.,   {Bessell} M.~S.,  1999,
  \mn@doi [\aj] {10.1086/301116}, \href
  {http://adsabs.harvard.edu/abs/1999AJ....118.2797K} {118, 2797}

\bibitem[\protect\citeauthoryear{{Koerwer}}{{Koerwer}}{2009}]{Koerwer09}
{Koerwer} J.~F.,  2009, \mn@doi [\aj] {10.1088/0004-6256/138/1/1}, \href
  {http://adsabs.harvard.edu/abs/2009AJ....138....1K} {138, 1}

\bibitem[\protect\citeauthoryear{{Kroupa}}{{Kroupa}}{2002}]{Kroupa02}
{Kroupa} P.,  2002, \mn@doi [Science] {10.1126/science.1067524}, \href
  {http://adsabs.harvard.edu/abs/2002Sci...295...82K} {295, 82}

\bibitem[\protect\citeauthoryear{{Lah}, {Kiss}  \& {Bedding}}{{Lah}
  et~al.}{2005}]{Lah05}
{Lah} P.,  {Kiss} L.~L.,   {Bedding} T.~R.,  2005, \mn@doi [\mnras]
  {10.1111/j.1745-3933.2005.00033.x}, \href
  {http://adsabs.harvard.edu/abs/2005MNRAS.359L..42L} {359, L42}

\bibitem[\protect\citeauthoryear{{Meschin}, {Gallart}, {Aparicio}, {Hidalgo},
  {Monelli}, {Stetson}  \& {Carrera}}{{Meschin} et~al.}{2014}]{meschin14}
{Meschin} I.,  {Gallart} C.,  {Aparicio} A.,  {Hidalgo} S.~L.,  {Monelli} M.,
  {Stetson} P.~B.,   {Carrera} R.,  2014, \mn@doi [\mnras]
  {10.1093/mnras/stt2220}, \href
  {http://adsabs.harvard.edu/abs/2014MNRAS.438.1067M} {438, 1067}

\bibitem[\protect\citeauthoryear{{Nikolaev}, {Drake}, {Keller}, {Cook},
  {Dalal}, {Griest}, {Welch}  \& {Kanbur}}{{Nikolaev}
  et~al.}{2004}]{Nikolaev04}
{Nikolaev} S.,  {Drake} A.~J.,  {Keller} S.~C.,  {Cook} K.~H.,  {Dalal} N.,
  {Griest} K.,  {Welch} D.~L.,   {Kanbur} S.~M.,  2004, \mn@doi [\apj]
  {10.1086/380439}, \href {http://adsabs.harvard.edu/abs/2004ApJ...601..260N}
  {601, 260}

\bibitem[\protect\citeauthoryear{{Olsen}}{{Olsen}}{1999}]{Olsen99}
{Olsen} K.~A.~G.,  1999, \mn@doi [\aj] {10.1086/300854}, \href
  {http://adsabs.harvard.edu/abs/1999AJ....117.2244O} {117, 2244}

\bibitem[\protect\citeauthoryear{{P{\'e}rez} et~al.,}{{P{\'e}rez}
  et~al.}{2017}]{Perez2017}
{P{\'e}rez} I.,  et~al., 2017, \mn@doi [\mnras] {10.1093/mnrasl/slx087}, \href
  {http://adsabs.harvard.edu/abs/2017MNRAS.470L.122P} {470, L122}

\bibitem[\protect\citeauthoryear{{Pietrinferni}, {Cassisi}, {Salaris}  \&
  {Castelli}}{{Pietrinferni} et~al.}{2004}]{Pietrinferni04}
{Pietrinferni} A.,  {Cassisi} S.,  {Salaris} M.,   {Castelli} F.,  2004,
  \mn@doi [\apj] {10.1086/422498}, \href
  {http://adsabs.harvard.edu/abs/2004ApJ...612..168P} {612, 168}

\bibitem[\protect\citeauthoryear{{Smecker-Hane}, {Cole}, {Gallagher}  \&
  {Stetson}}{{Smecker-Hane} et~al.}{2002}]{smeckerhane02}
{Smecker-Hane} T.~A.,  {Cole} A.~A.,  {Gallagher} III J.~S.,   {Stetson} P.~B.,
   2002, \mn@doi [\apj] {10.1086/337985}, \href
  {http://adsabs.harvard.edu/abs/2002ApJ...566..239S} {566, 239}

\bibitem[\protect\citeauthoryear{{Staveley-Smith}, {Kim}, {Calabretta},
  {Haynes}  \& {Kesteven}}{{Staveley-Smith} et~al.}{2003}]{Staveley03}
{Staveley-Smith} L.,  {Kim} S.,  {Calabretta} M.~R.,  {Haynes} R.~F.,
  {Kesteven} M.~J.,  2003, \mn@doi [\mnras] {10.1046/j.1365-8711.2003.06146.x},
  \href {http://adsabs.harvard.edu/abs/2003MNRAS.339...87S} {339, 87}

\bibitem[\protect\citeauthoryear{{Stetson}}{{Stetson}}{1987}]{stetson87}
{Stetson} P.~B.,  1987, \mn@doi [\pasp] {10.1086/131977}, \href
  {http://adsabs.harvard.edu/abs/1987PASP...99..191S} {99, 191}

\bibitem[\protect\citeauthoryear{{Stetson}}{{Stetson}}{1994}]{stetson94}
{Stetson} P.~B.,  1994, \mn@doi [\pasp] {10.1086/133378}, \href
  {http://adsabs.harvard.edu/abs/1994PASP..106..250S} {106, 250}

\bibitem[\protect\citeauthoryear{{Stinson}, {Dalcanton}, {Quinn}, {Gogarten},
  {Kaufmann}  \& {Wadsley}}{{Stinson} et~al.}{2009}]{stinson09}
{Stinson} G.~S.,  {Dalcanton} J.~J.,  {Quinn} T.,  {Gogarten} S.~M.,
  {Kaufmann} T.,   {Wadsley} J.,  2009, \mn@doi [\mnras]
  {10.1111/j.1365-2966.2009.14555.x}, \href
  {http://adsabs.harvard.edu/abs/2009MNRAS.395.1455S} {395, 1455}

\bibitem[\protect\citeauthoryear{{Subramaniam}}{{Subramaniam}}{2003}]{Subramaniam03}
{Subramaniam} A.,  2003, Bulletin of the Astronomical Society of India, \href
  {http://adsabs.harvard.edu/abs/2003BASI...31..413S} {31, 413}

\bibitem[\protect\citeauthoryear{{Subramaniam} \& {Subramanian}}{{Subramaniam}
  \& {Subramanian}}{2009}]{Subramaniam09}
{Subramaniam} A.,  {Subramanian} S.,  2009, \mn@doi [\apjl]
  {10.1088/0004-637X/703/1/L37}, \href
  {http://adsabs.harvard.edu/abs/2009ApJ...703L..37S} {703, L37}

\bibitem[\protect\citeauthoryear{{Weisz}, {Dolphin}, {Skillman}, {Holtzman},
  {Dalcanton}, {Cole}  \& {Neary}}{{Weisz} et~al.}{2013}]{Weisz13}
{Weisz} D.~R.,  {Dolphin} A.~E.,  {Skillman} E.~D.,  {Holtzman} J.,
  {Dalcanton} J.~J.,  {Cole} A.~A.,   {Neary} K.,  2013, \mn@doi [\mnras]
  {10.1093/mnras/stt165}, \href
  {http://adsabs.harvard.edu/abs/2013MNRAS.431..364W} {431, 364}

\bibitem[\protect\citeauthoryear{{Wilcots}}{{Wilcots}}{2008}]{Wilcots08}
{Wilcots} E.~M.,  2008, \mn@doi [Astrophysics and Space Science Proceedings]
  {10.1007/978-1-4020-6933-8_14}, \href
  {http://adsabs.harvard.edu/abs/2008ASSP....5...69W} {5, 69}

\bibitem[\protect\citeauthoryear{{Zaritsky}}{{Zaritsky}}{2004}]{Zaritsky04}
{Zaritsky} D.,  2004, \mn@doi [\apjl] {10.1086/425312}, \href
  {http://adsabs.harvard.edu/abs/2004ApJ...614L..37Z} {614, L37}

\bibitem[\protect\citeauthoryear{{Zhao} \& {Evans}}{{Zhao} \&
  {Evans}}{2000}]{Zhao00}
{Zhao} H.,  {Evans} N.~W.,  2000, \mn@doi [\apjl] {10.1086/317324}, \href
  {http://adsabs.harvard.edu/abs/2000ApJ...545L..35Z} {545, L35}

\bibitem[\protect\citeauthoryear{{de Vaucouleurs} \& {Freeman}}{{de
  Vaucouleurs} \& {Freeman}}{1972}]{Vaucouleurs72}
{de Vaucouleurs} G.,  {Freeman} K.~C.,  1972, \mn@doi [Vistas in Astronomy]
  {10.1016/0083-6656(72)90026-8}, \href
  {http://adsabs.harvard.edu/abs/1972VA.....14..163D} {14, 163}

\bibitem[\protect\citeauthoryear{{van der Marel}}{{van der
  Marel}}{2001}]{marel01}
{van der Marel} R.~P.,  2001, \mn@doi [\aj] {10.1086/323100}, \href
  {http://adsabs.harvard.edu/abs/2001AJ....122.1827V} {122, 1827}

\makeatother
\end{thebibliography}





\appendix


\bsp	
\label{lastpage}
\end{document}